\documentstyle[12pt,fleqn]{article}

\font\twlgot =eufm10 scaled \magstep1
\font\egtgot =eufm8
\font\sevgot =eufm7
\font\twlmsb =msbm10 scaled \magstep1
\font\egtmsb =msbm8
\font\sevmsb =msbm7

\newfam\gotfam

\textfont\gotfam\twlgot
\scriptfont\gotfam\egtgot
\scriptscriptfont\gotfam\sevgot

\newfam\msbfam
\textfont\msbfam\twlmsb
\scriptfont\msbfam\egtmsb
\scriptscriptfont\msbfam\sevmsb
\def\Bbb{\protect\pBbb}
\def\pBbb{\relax\ifmmode\expandafter\Bb\else\typeout{You cann't use
Bbb in text mode}\fi}
\def\Bb #1{{\fam\msbfam\relax#1}}

\def\thebibliography#1{\section*{References}\list
  {[\arabic{enumi}]}{\settowidth\labelwidth{#1}\leftmargin\labelwidth
    \advance\leftmargin\labelsep
    \usecounter{enumi}}
    \def\newblock{\hskip .11em plus .33em minus .07em}
    \sloppy\clubpenalty4000\widowpenalty4000
    \sfcode`\.=1000\relax}

\def\op#1{\mathop{\fam0 #1}\limits}

\newcommand{\beq}{\begin{equation}}
\newcommand{\eeq}{\end{equation}}
\newcommand{\ben}{\begin{eqnarray}}
\newcommand{\een}{\end{eqnarray}}
\newcommand{\be}{\begin{eqnarray*}}
\newcommand{\ee}{\end{eqnarray*}}
\newcommand{\bea}{\begin{eqalph}}
\newcommand{\eea}{\end{eqalph}}

\newcommand{\cF}{{\cal F}}

\newcommand{\bL}{{\bf L}}

\newcommand{\al}{\alpha}

\newcommand{\bt}{\beta}
\newcommand{\dl}{\delta}
\newcommand{\la}{\lambda}

\newcommand{\m}{\mu}

\newcommand{\g}{\gamma}

\newcommand{\ol}{\overline}
\newcommand{\dr}{\partial}

\newcommand{\ve}{\varepsilon}

\newcounter{eqalph}
\newcounter{equationa}
\newcounter{theorem}
\newcounter{remark}
\newcounter{proposition}
\newcounter{lemma}
\newcounter{corollary}
\newcounter{definition}

\newenvironment{eqalph}{\stepcounter{equation}
\setcounter{equationa}{\value{equation}}
\setcounter{equation}{0}

\begin{eqnarray}}{\end{eqnarray}\setcounter{equation}{\value{equationa}}}

\def\theremark{\arabic{remark}}

\def\thedefinition{\arabic{definition}}

\newenvironment{proof}{\noindent 
{\bf Proof.}}{\hfill $\Box$ \medskip}

\newenvironment{prop}{\refstepcounter{definition} 
\medskip\noindent{\bf Proposition \thedefinition.}\it }{\medskip}

\newcommand{\mar}[1]{}

\begin{document}
\hbox{}

{\parindent=0pt

{\large\bf The covariant Lyapunov tensor and the Lyapunov stability
with respect to time-dependent Riemannian metrics}
\bigskip

{\sc  Gennadi 
Sardanashvily$\dagger$}
\bigskip

\begin{small}
$\dagger$ Department of Theoretical Physics, Physics Faculty, Moscow State
University, 117234 Moscow, Russia; E-mail: sard@grav.phys.msu.su 

\bigskip


{\bf Abstract.}
We show that any solution of a 
smooth first order dynamic equation can be made
Lyapunov stable at will by the choice of an
appropriate time-dependent Riemannian metric.
\end{small}
}


\section{Introduction}

The Lyapunov matrix of a first order dynamic equation 
is defined as the coefficient matrix of 
the variation equation, and its properties are used as 
a criteria of stability of solutions of this equation
\cite{gal,hirs}. The problem is that the Lyapunov matrix
is not a tensor under 
coordinate transformations, unless they are linear
and time-independent. 
For instance, any
first order dynamic equation defines a coordinate atlas
such that its solution is constant on each coordinate chart 
(see Proposition \ref{ch6} below) and, consequently, the Lyapunov matrix 
vanishes.

We introduce the covariant Lyapunov tensor of a first order
dynamic equation and study the Lyapunov stability of
solutions when this tensor is 
negative-definite. The covariant Lyapunov tensor
essentially depends on the choice of a Riemannian metric. We show that, if a
dynamic equation is defined by a complete vector field, the Lyapunov
exponent of its solution can be made equal to any real number
with respect to the appropriate
time-dependent Riemannian metric. It follows that
chaos in dynamical systems described by smooth ($C^\infty$) first
order dynamic equations can be characterized in full by time-dependent
Riemannian metrics.

\section{Geometry of first order dynamic equations 
in non-autonomous mechanics}

Let $\Bbb R$ be the time axis provided with the Cartesian coordinate $t$ 
and transition functions $t'=t+$const.
In geometric terms, a (smooth) first order dynamic equation in non-autonomous
mechanics is defined as 
a vector field $\g$ on a smooth fibre bundle 
\mar{ch0}\beq
\pi:Y\to\Bbb R \label{ch0}
\eeq
which obeys the condition 
$\g\rfloor dt=1$ \cite{book98}, i.e.,
\mar{ch3}\beq
\g=\dr_t + \g^\la\dr_\la. \label{ch3}
\eeq
The associated first order dynamic
equation takes the form
\mar{ch5}\beq
\dot t=1, \qquad \dot y^\la=\g^\la(t,y^\m)\dr_\la, \label{ch5}
\eeq
where $(t,y^\la,\dot t,\dot y^\la)$ are holonomic coordinates on $TY$.
Its solutions are trajectories of the vector field
$\g$ (\ref{ch3}). They assemble into 
a (regular) foliation $\cF$ of $Y$. Equivalently, $\g$ (\ref{ch3})
is defined as a connection on the fibre bundle (\ref{ch0}).

A fibre bundle $Y$ (\ref{ch0}) is trivial, but it admits different 
trivializations  
\mar{ch4}\beq
Y\cong\Bbb R\times M, \label{ch4}
\eeq
distinguished by fibrations $Y\to M$. Moreover, Proposition \ref{ch6} below
appeals to an atlas of $Y$ as a fibred manifold.
If there is a trivialization (\ref{ch4})
such that,
with respect to the associated coordinates, 
the components
$\g^\la$ of the connection $\g$ (\ref{ch3}) are independent of $t$,
one says that $\g$ is a conservative first order dynamic equation
on $M$.

\begin{prop} \label{ch6} \mar{ch6}
Given a first order dynamic equation $\g$, there exists an 
atlas $\Psi=\{(U;t,y^a)\}$ 
of a fibred manifold $Y\to\Bbb R$ with time-independent 
transition functions $y'^a(y^b)$
such that any solution $s$ of $\g$ on each chart $(U;t,y^a)$ reads
\be
s^a(t)={\rm const}., \qquad t\in \pi(U)\subset \Bbb R.
\ee
\end{prop}

\begin{proof}
The atlas $\Psi$ is an atlas of adapted coordinates for the 
foliation $\cF$ of trajectories of the vector field (\ref{ch3})
\cite{rei}. 
\end{proof}

\begin{prop} \label{ch8} \mar{ch8}
Let the vector field $\g$ (\ref{ch3}) be complete,
i.e., there is a unique global 
solution of the dynamic equation $\g$ through each point of $Y$.
Then there is a trivialization
(\ref{ch4}) of $Y$ such that any solution $s$ of $\g$ reads
\be
s^a(t)={\rm const}., \qquad t\in \Bbb R, 
\ee
with respect to associated bundle coordinates $(t,y^a)$.
\end{prop}

\begin{proof}
If $\g$ is complete, the foliation $\cF$ of its trajectories
is a fibration of $Y$ along these trajectories onto any fibre of $Y$, e.g.,
 $Y_{t=0}\cong M$. This fibration yields a desired trivialization
\cite{book98,book00}.
\end{proof}

 One can think of the coordinates $(t,y^a)$ in Proposition \ref{ch8} as being 
the initial date coordinates because all points of the same 
trajectory differ from each other only in the temporal coordinate. 

Let us consider the canonical lift $V\g$ of the vector 
field $\g$ (\ref{ch3}) onto the vertical tangent bundle $VY$ of $Y\to\Bbb R$. 
With respect to the holonomic bundle coordinates $(t,y^\la,\ol y^\la)$
on $VY$, it reads
\mar{ch11}\beq
V\g= \g  +
\dr_\m\g^\la \ol y^\m\ol\dr_\la, \qquad \ol\dr_\la=\frac{\dr}{\dr \ol y^\la}.
\label{ch11}
\eeq
This vector field obeys the condition $V\g\rfloor dt=1$, and 
 defines the first order dynamic
equation 
\mar{ch12}\bea
&& \dot t=1, \qquad \dot y^\la=\g^\la(t,y^\nu), \label{ch12a} \\
&& \dot {\ol y_t}^\la= 
\dr_\m\g^\la(t,y^\nu)\ol y^\m\label{ch12b}
\eea
on $VY$. The equation (\ref{ch12a}) coincides with the initial one (\ref{ch5}).
The equation (\ref{ch12b}) is the well-known variation
equation. 
Substituting a solution $s$ of the initial dynamic equation
(\ref{ch12a}) into (\ref{ch12b}), one obtains a linear dynamic equation
whose  solutions $\ol s$ are Jacobi fields of
the solution $s$. In particular, if $Y\to\Bbb R$ is
a vector bundle, there is the canonical splitting $VY\cong Y\times Y$ and
the morphism $VY\to Y$ so that 
$s +\ol s$ obeys the initial dynamic
equation (\ref{ch12a}) modulo the terms of order $>1$ in $\ol s$.

Remark that, if $\g$ is the Hamilton equation for a Hamiltonian form $H$
on 
\be
Y=V^*Q\to Q\to\Bbb R
\ee
in time-dependent
Hamiltonian mechanics, the variation equation $V\g$ is also the Hamilton 
equation for the Hamiltonian form on $V^*VQ$ which is the canonical tangent lift
of $H$ onto $VV^*Q=V^*VQ$ \cite{gia99,book98,book00}. 

\section{The covariant Lyapunov tensor}

The collection of coefficients 
\mar{ch70}\beq
l_\m{}^\la=\dr_\m\g^\la \label{ch70}
\eeq
of the variation equation (\ref{ch12b}) is called the Laypunov matrix.
Clearly, it is not a tensor under
bundle coordinate transformations of the fibre bundle $Y$ (\ref{ch0}).
Let us bring $l_\m{}^\la$ into a covariant tensor.

The stability conditions of Lyapunov for solutions of a first order
dynamic equation involve the notion of a distance between
different solutions at an instant $t$. Therefore, let 
a fibre bundle $Y\to\Bbb R$ be provided with a Riemannian fibre metric
$g$, defined as a section of the symmetrized tensor product 
$\op\vee^2 V^*Y\to Y$ of the
vertical cotangent bundle $V^*Y$ of $Y\to\Bbb R$. With respect to the
holonomic coordinates $(t,y^\la,\ol y_\la)$ on $V^*Y$,
it takes the coordinate form
\mar{ch15}\beq
g=\frac12 g_{\al\bt}(t,y^\la)\ol dy^\al\vee \ol dy^\bt, \label{ch15}
\eeq  
where $\{\ol dy^\m\}$ are the holonomic fibre bases for $V^*Y$.

Given a first order differential equation $\g$, let 
\mar{ch13}\beq
V^*\g= \g -
\dr_\m\g^\la \ol y_\la\ol\dr^\m, \qquad \ol\dr^\m=\frac{\dr}{\dr \ol y_\m}.
\label{ch13}
\eeq
be the canonical lift of the vector field $\g$ (\ref{ch3}) onto $V^*Y$.
It is a connection on $V^*Y\to\Bbb R$. Let us consider the Lie derivative
$\bL_\g g$ of the Riemannian fibre metric $g$ along the vector 
field $V\g$ (\ref{ch13}). It reads
\mar{ch14}\beq
L_{\al\bt}=(D_tg)_{\al\bt}= \dr_t g_{\al\bt}+ \g^\la\dr_\la g_{\al\la}
+ \dr_\al\g^\la g_{\la\bt} + \dr_\bt\g^\la g_{\al\la}. \label{ch14}
\eeq
This is a tensor with respect 
to any bundle coordinate
transformation of the fibre bundle (\ref{ch0}).
We agree to call it the covariant Lyapunov tensor. 
If $g$ is an Euclidean metric, it comes to symmetrization
\be
 L_{\al\bt}=\dr_\al\g^\bt + \dr_\bt\g^\al=l_\al{}^\bt +l_\bt{}^\al
\ee
of the Lyapunov matrix (\ref{ch70}).

Let us point the following two properties of the covariant 
Lyapunov tensor.

(i) Written with respect to the atlas $\Psi$ in Proposition \ref{ch6},
the covariant Lyapunov tensor is
\mar{ch72}\beq
L_{ab}= \dr_t g_{ab}. \label{ch72}
\eeq

(ii) Given a solution $s$
of the dynamic equation $\g$ and 
a solution $\ol s$ of the variation equation (\ref{ch12b}), we have 
\mar{ch71}\beq
L_{\al\bt}(t,s^\la(t))\ol s^\al\ol s^\bt=
\frac{d}{dt}(g_{\al\bt}(t,s^\la(t))\ol s^\al\ol s^\bt). 
\label{ch71}
\eeq

The definition of the covariant Lyapunov tensor (\ref{ch14})
depends on the choice of a Riemannian fibre metric on the
fibre bundle $Y$. 

\begin{prop} \label{ch73} \mar{ch73}
If the vector field $\g$ is complete, there is a Riemannian fibre metric
on $Y$ such that the covariant Lyapunov tensor vanishes everywhere.
\end{prop}

\begin{proof}
Let us choose the atlas of the initial date coordinates in 
Proposition \ref{ch8}. Using the fibration $Y\to Y_{t=0}$,
one can provide $Y$ with a time-independent
Riemannian fibre metric 
\mar{ch80}\beq
g_{ab}(t,y^a)=h(t)g^0_{ab}(y^a) \label{ch80}
\eeq
 where
$g^0_{ab}(y^a)$ is a Riemannian metric on the fibre $Y_{t=0}$ and $h(t)$ 
is a positive smooth function on $\Bbb R$. The covariant 
Lyapunov tensor with respect to the metric (\ref{ch80})
is
\be
L_{ab}=\dr_t h g_{ab}.
\ee
Putting $h(t)=1$, we obtain $L=0$.
\end{proof}

\section{The local Lyapunov stability} 

With the covariant Lyapunov tensor (\ref{ch14}), we obtain the
following variant of the stability condition of Lyapunov.

By the instantwise distance $\rho_t(s,s')$ 
between two solutions $s$ and $s'$ 
of the dynamic equation $\g$ at an instant $t$
is meant the distance between the points $s(t)$ and $s'(t)$
in the Riemannian space $(Y_t,g(t))$.
Recall that a solution $s$ of a first order dynamic 
equation $g$
is said to be locally Lyapunov stable at an instant
$t\in\Bbb R$ if, for
any $\ve>0$, 
there is $\dl>0$ such that $\rho_t(s,s')<\dl$ implies 
$\rho_{t'}(s,s')<\ve$ for each $t'$ which belongs to
some half closed interval $[t,.)$. Being locally Lyapunov stable
with respect to $g$, a solution is so
with respect to any Riemannian fibre metric on $Y$.

\begin{prop} \label{ch16} \mar{ch16}
Let the covariant Lyapunov tensor $L$ (\ref{ch14}) at a point
$y\in Y$ be a negative-definite bilinear form
on $V_yY$. Then there is an open neighbourhood $U_y$ of 
$y$ and an open tubular neighbourhood $U_s\subset U_y$ of the 
trajectory $s$ through $y$ in $U_y$ such that 
\mar{ch17}\beq 
\rho_{t'>t}(s,s')<\rho_t(s,s')
\label{ch17}
\eeq 
for any $t'\in [t,)\subset \pi(U_s)$ and $s'$ crossing $U_s$. 
\end{prop}

\begin{proof}
Since the condition and the statement of Proposition \ref{ch16}
are coordinate-independent, let us choose a chart $(U;t,y^a)$
of the bundle atlas $\Psi$ in Proposition \ref{ch6} which 
cover the point $y$. With respect to these coordinates, the covariant
Lyapunov tensor takes the form (\ref{ch72}).
Restricted to 
$U$, the leaf $s$ of the foliation $\cF$ 
is an imbedded submanifold, and has an open tubular 
neighbourhood $W$, i.e., any trajectory in $U$
intersecting $W$ does not leave $W$.
There is an open neighbourhood $U_y\subset U$ of 
the point $y\in Y$
where the Lyapunov tensor $L$ (\ref{ch14}) holds negative-definite.
Put $W'=W\cap U_y$.
The foliation $\cF$ restricted to the tubular $W'$ defines its 
fibration 
\be
\zeta: W'\to W'\cap Y_{\pi(y)}
\ee
and the corresponding trivialization
\mar{ch35}\beq
W'\cong \pi(W')\times (W'\cap Y_{\pi(y)}). \label{ch35}
\eeq
There is an open neighbourhood $V\subset 
Y_{\pi(y)}\cap W'$ of $y$ in $Y_{\pi(y)}$ which can be provided
with the normal coordinates $(x^a)$ defined by  
the Riemannian metric
$g$ in $Y_{\pi(y)}$ and centralized at $y$. Let us 
consider the open tubular $U_s=\zeta^{-1}(V)$. It is a subbundle
of the trivial bundle (\ref{ch35}) endowed with the
coordinates $(t,x^a)$, $t\in(,)$. Without a loss of generality,
put $t=\pi(y)=0$.
 With respect to these coordinates, the solution
$s$ in $U_s$ reads $s^a(t)=0$. Let 
\mar{ch30}\beq
s'^a(t)=u^a={\rm const}., \qquad t\in (,)=\pi(U_s), \label{ch30}
\eeq
be another solution crossing $U_s$.
The instantwise distance $\rho_t(s,s')$ between solutions $s$ and
$s'$ is the distance between the points
$(t,0)$ and $(t,u)$ in the Riemannian space $(Y_t,g(t))$. 
This distance does not
exceed the length 
\mar{ch31}\beq
\ol\rho_t(s,s')=\left[\op\int^1_0 g_{ab}(t,\tau u^c)
u^a u^b d\tau\right]^{1/2}
\label{ch31}
\eeq
of the curve 
\mar{ch34}\beq
x^a=\tau u^a, \qquad \tau\in [0,1] \label{ch34}
\eeq
in the Riemannian space $(Y_t,g(t))$. At the same time, we have
\be
\rho_{t=0}(s,s')=\ol \rho_{t=0}(s,s')
\ee
The temporal derivative of the function
$\ol\rho_t(s,s')$ (\ref{ch31}) reads
\mar{ch32}\beq
\dr_t\ol\rho_t(s,s')= \frac{1}{2(\ol\rho_t(s,s'))^{1/2}}
\op\int^1_0 \dr_tg_{ab}(t,\tau u^c)
u^a u^b d\tau. \label{ch32}
\eeq
Since the bilinear form $\dr_tg_{ab}=L_{ab}$ is negative-definite
at all points of the curve (\ref{ch34}), the derivative 
(\ref{ch32}) at all points $t\in (,)$ is also negative. Hence,
we obtain 
\be
\rho_{t'>0}(s,s')<\ol \rho_{t'>0}(s,s')<\ol \rho_0(s,s')= \rho_0(s,s').
\ee
\end{proof}

The inequality (\ref{ch17}) shows that a solution of a first order
differential equation which obeys the condition of Proposition \ref{ch16}
is locally Lyapunov stable at an instant $t$. 
Moreover, one can say that it is isometrically stable at $t$, and is so
in some neighbourhood of $t$. Given another Riemannian fibre metric on
$Y$, this solution remains locally Lyapunov stable, but need not
be isometrically stable.

\section{The asymptotic Lyapunov stability}

Let $\g$ (\ref{ch3}) be a complete vector field. With a minor modification 
of the proof of Proposition \ref{ch16}, one can state the following.

\begin{prop} \label{ch40} \mar{ch40}
Let $s$ be a solution
of the first order dynamic equation $\g$. If there exists 
an open tubular neighbourhood $U_s$ of the trajectory $s$
where the covariant Lyapunov tensor
(\ref{ch14}) is negative-definite at all instants $t$ exceeding some
$t_0$, then 
\mar{ch50}\beq
\lim_{t'\to\infty}[\rho_{t'}(s,s')-\rho_t(s,s')]<0 
\label{ch50}
\eeq
for any $t>t_0$ and any solution $s'\subset U_s$.
\end{prop}

One can say that a solution $s$ in Proposition \ref{ch40} 
obeys the isometric asymptotic stability of Lyapunov.
Of course, it is also asymptotically stable, but so is any
solution of a smooth dynamic equation as follows. 

\begin{prop} \label{ch75} \mar{ch75}
Any solution of any first order dynamic equation defined by
a complete vector field (\ref{ch3}) is Lyapunov stable
at each instant of time and is asymptotically 
Lyapunov stable.
\end{prop}

\begin{proof}
These properties obviously hold with respect to
the Riemannian fibre metric (\ref{ch80}) in Proposition 
\ref{ch73} where $h=1$,
and so do with respect to any Riemannian fibre
metric on $Y$.
\end{proof}

It follows that a solution of a first order dynamic equation
is Lyapunov unstable only if this equation is given by a 
non-complete vector field $\g$ (\ref{ch3}). 
In this case, there exists a local
(but not necessarily global) Riemannian fibre metric on $Y$
such that a solution $s$ of $\g$ is locally stable 
at a given instant $t$.

One can improve Proposition \ref{ch75} as follows.

\begin{prop} \label{ch90} \mar{ch90}
Let $\la$ be a real number. Given a dynamic equation $\g$
defined by a complete vector field $\g$ (\ref{ch3}) and
its solution $s$,
there is a Riemannian fibre metric on $Y$ such that
the Lyapunov spectrum reduces to  $\la$.
\end{prop}

\begin{proof}
Recall that the (upper) Lyapunov exponent of a 
solution $s'$ with respect to
a solution $s$ is defined as the limit 
\mar{ch51'}\beq
K(s,s')=\lim^-_{t\to\infty} \frac{1}{t}
\ln(\rho_t(s,s')). \label{ch51'}
\eeq
Let provide the fibre bundle $Y$ with the 
Riemannian fibre metric (\ref{ch80}) in Proposition
\ref{ch73} where $h=\exp(\la t)$. A simple computation shows
that the Laypunov exponent (\ref{ch51'}) with
respect to this metric is exactly $\la$.
\end{proof}

If the upper limit
\be
\lim^-_{\rho_{t=0}(s,s')\to 0}K(s,s')=\la
\ee
is negative, the solution $s$ is said to be exponentially Lyapunov
stable. If there exists at least one positive Lyapunov exponent, one speaks 
on chaos in a dynamical system \cite{gutz}.
Proposition \ref{ch90} shows that chaos in smooth dynamical systems 
can be characterized in full by time-dependent Riemannian metrics.

\end{document}